\pdfoutput=1
\documentclass[12pt,a4paper]{article}
\usepackage{xcolor}
\usepackage{enumerate}
\usepackage{framed}
\usepackage{mdframed}
\usepackage{amsmath}
\usepackage{amsfonts}
\usepackage{mathrsfs}
\usepackage{amssymb}
\usepackage{graphicx, rotating}
\usepackage{epsfig}
\usepackage{latexsym}
\usepackage{graphicx}
\usepackage{color}
\usepackage{amsmath,bm,amssymb}
\usepackage{cite}
\usepackage{slashed}
\usepackage{hyperref}
\usepackage{epstopdf}
\usepackage[font=footnotesize,labelfont=]{caption}
\AppendGraphicsExtensions{.tif}
\hypersetup{colorlinks=true, citecolor=bluscuro, linkcolor=black, urlcolor=bluscuro}
\definecolor{rossos}{cmyk}{0,1,1,0.55}
\definecolor{bluscuro}{rgb}{0.15, 0.2, .85}
\definecolor{bluchiaro}{cmyk}{1,.3,0.,0.1}

\def\0{\vec{0}}

\newcommand{\hk}{\hat{k}}

\def\vx{{\vec{x}}}
\def\vz{{\vec{z}}}

\def\vk{{\vec{k}}}

\def\beq{\begin{equation}}
\def\eeq{\end{equation}}

\newcommand{\D}{\Delta}

\begin{document}
\def\thefootnote{\fnsymbol{footnote}}

\begin{center}
\Large{\textbf{Imprints of Spinning Particles \\ on Primordial Cosmological Perturbations }} \\[0.5cm]
\end{center}
\vspace{0.5cm}

\begin{center}

\large{Gabriele Franciolini$^{\rm a}$\footnote{gabriele.franciolini@unige.ch},  Alex Kehagias$^{\rm b}\footnote{kehagias@central.ntua.gr}$  and  Antonio Riotto$^{\rm a}$ \footnote{antonio.riotto@unige.ch}}
\\[0.5cm]

\small{
\textit{$^{\rm a}$Department of Theoretical Physics and Center for Astroparticle Physics (CAP) \\
24 quai E. Ansermet, CH-1211 Geneva 4, Switzerland}}

\small{
\textit{$^{\rm b}$Physics Division, National Technical University of Athens, 15780 Zografou Campus, Athens, Greece}}

\vspace{.2cm}


\vspace{.2cm}

\vspace{.2cm}

\end{center}

\vspace{.7cm}

\hrule \vspace{0.3cm}
\noindent \small{\textbf{Abstract}\\ 
If there exist higher-spin  particles during inflation which are light compared to the Hubble rate, they may leave
distinct statistical anisotropic imprints on the  correlators involving  scalar and graviton  fluctuations. 
We characterise such signatures using the dS/CFT$_3$ correspondence and the operator product expansion techniques. In particular,  we obtain
generic results for the case of partially massless higher-spin states. }

\vspace{0.3cm}
\noindent
\hrule
\def\thefootnote{\arabic{footnote}}
\setcounter{footnote}{0}

%
%
%
%


 \def\vx{\vec{ x}} 
\def\vk{\vec{k}}
\def\vy{\vec{y}}

\numberwithin{equation}{section}

\def\la{~\mbox{\raisebox{-.6ex}{$\stackrel{<}{\sim}$}}~}
\def\ga{~\mbox{\raisebox{-.6ex}{$\stackrel{>}{\sim}$}}~}
\def\bq{\begin{quote}}
\def\eq{\end{quote}}
\def\PL{{ \it Phys. Lett.} }
\def\PRL{{\it Phys. Rev. Lett.} }
\def\NP{{\it Nucl. Phys.} }
\def\PR{{\it Phys. Rev.} }
\def\MPL{{\it Mod. Phys. Lett.} }
\def\IJMP{{\it Int. J. Mod .Phys.} }
\font\tinynk=cmr6 at 10truept
\newcommand{\be}{\begin{eqnarray}}
\newcommand{\ee}{\end{eqnarray}}
\newcommand{\n}{{\bf n}}
\newcommand{\arXiv}[2]{\href{http://arxiv.org/pdf/#1}{{\tt [#2/#1]}}}
\newcommand{\arXivold}[1]{\href{http://arxiv.org/pdf/#1}{{\tt [#1]}}}

\section{Introduction \label{sec:intro}} 
Up to now, the most robust and successful mechanism to explain the primordial seeds for the cosmic microwave background anisotropies and the    large-scale structure we observe in the universe  is inflation \cite{inf}, that is a (quasi-)de Sitter period  when the physical space expands almost exponentially 
and quantum fluctuations initially at microscopic scales are stretched to macroscopical  scales. After   horizon re-entry,     they  initiate the phenomenon of  gravitational instability   giving rise to the structures of the  universe. 

From the high energy point of view, inflation is an appealing playground as it may happen at energies much larger than the electroweak scale and thus
provide the most powerful collider to test physics at high energy \cite{chen1,chen2,am}. For instance,   cosmological correlators of the comoving curvature perturbation   may be non-gaussian  and originated from the exchange of massive higher-spin  fields \cite{am,baumann,chen,azadeh}. This  generates some hope to  learn something  about their masses and    spins.

A step towards the general characterisation of the cosmological perturbations generated during a de Sitter epoch has been the formulation of the so-called
dS/CFT$_3$ correspondence \cite{strominger}. During a period of exact exponentially expansion, the isometries of the corresponding dS spacetime form 
a  SO(1,4) group which    acts  as the conformal group of a  CFT$_3$ on 
 $\mathbb{R}^3$ and    on the super-Hubble perturbations. 

Technically,  a four-dimensional  field $A_{\mu_1\cdots\mu_s}(\vx,\tau)$ with mass $m$ and spin $s$
evolves such that $ A_{i_1\cdots i_s}(\vx,\tau) =(-\tau)^{\Delta-s} A_{i_1\cdots i_s}(\vx)$ when approaching the 
boundary $\tau=0$ ($\tau$ is the conformal time) with
\be
\label{aa}
\Delta=\frac{3}{2}-\sqrt{\left(s-\frac{1}{2}\right)^2 -\frac{m^2}{H^2}},
\ee
$H$ being  the Hubble rate during inflation.
The field $A_{i_1\cdots i_s}(\vx)$  behaves like a primary field with conformal weight $\Delta$ under the  boundary conformal transformations. At this stage, it is interesting to point out that one can give two different interpretations of the SO(1,4) group. When gravitational fluctuations are excited, a case relevant for single-field models of inflation, SO(1,4) is  identified with the three-dimensional conformal group at different constant  time-slices; in the limit in which gravity is decoupled, holding when one is interested in spectator fields (such as in the curvaton model \cite{curvaton}), including higher-spin fields, SO(1,4) is a non-linearly realised symmetry of the action in de Sitter. 

De Sitter isometries play an important role when dealing with massive  spinning fields: the mass $m$ and the spin $s$ of a  higher-spin field  in de Sitter must respect the   Higuchi bound \cite{hig} to avoid the lower helicity modes  to become ghost-like
\beq
m^2>s(s-1)H^2,
\eeq
This inequality can be inferred  from the dS/CFT$_3$ correspondence \cite{am,higtensor,higper} and  implies that higher-spin super-Hubble  two-point correlators $\langle A_{i_1\cdots i_s}A^{i_1\cdots i_s}\rangle$  decay    as  $(-\tau)^{2\Delta}$ with $\Delta>1$, as one can deduce from the relation (\ref{aa}).
Higher-spin fields are therefore  short-lived and their impact on cosmological observables is rather suppressed. 

To the best of our knowledge there are two ways to evade the Higuchi bound.  
On one side, one can exploit partially massless higher-spin fields \cite{p1,p2,p3,p4,p5}. Indeed, for some particular values of their masses,  some helicities of spinning states may acquire a vanishing conformal weight in de Sitter so that their fluctuations can be excited with a scale-invariant spectrum during inflation. From the point of view of the dS/CFT$_3$ correspondence, these states  correspond to   rank-$s$ symmetric
boundary tensors    which are partially conserved \cite{witten,dw}.   

On the other side, to obtain vanishing conformal weights for  massive  higher-spin fields, one can couple the higher-spin states to a preferred foliation of spacetime such that the quadratic action is not covariant.  This phenomenon may take place  by coupling the higher-spin fields   to a suitable function of time (or better of the classical value $\phi_0$ of the inflaton field). For instance, for  spin-1 fluctuations a modification of the kinetic term of the form  $I(\phi_0)F_{\mu\nu}^2$ may lead to 
constant super-Hubble perturbations of the electric or magnetic fields  \cite{v1,v2,muk} and the scalar and vector correlators can be determined exploiting the dS/CFT$_3$ correspondence \cite{vector}. An extension of such  mechanism to higher-spin 
has been done in Ref. 
\cite{noi}. There  suitable  time-dependent functions of time  coupled to the higher-spin fields have been identified in such a way that the lower helicity modes  are prevented from becoming ghost-like, still allowing masses below   the Higuchi bound and preserving the correct number of degrees of freedom for the higher-spin fields. These couplings   give rise to enhanced symmetries and  correlators decaying  slower than what dictated by the Higuchi bound outside the Hubble radius.

When spinning degrees of freedom  are quantum mechanically excited and remain constant on super-Hubble scales, they leave  a distinctive statistical anisotropic signature  on the cosmological  correlators \cite{watanabe,soda,peloso,noianis}.  This happens because during the prolonged period of inflation  an infrared background of the higher-spin field is generated with a magnitude of order of the square root of its variance. During the last 60 e-folds or so, 
when the scalar modes exit the Hubble radius
at   comoving wavenumbers corresponding  to cosmologically relevant length scales, and 
 taking our observed universe as a single realisation of the different ensemble ones \cite{peloso},
 they live in a background which is slightly anisotropic due to such non-vanishing  infrared vacuum expectation value of the spinning fields.  The statistical anisotropy is distinctive since  the angle structure depends  on the
spins. If observed, the induced  anisotropies  will  deliver  fundamental  informations about the particle content at high energies.

The set-up we will be considering therefore is the following, see Fig. 1:
\begin{enumerate}
\item there exist spinning light degrees of freedom whose two-point correlators are approximately constant on super-Hubble scales; they provide a  representation of de Sitter isometry group and coincide with those of the CFT$_3$ on super-Hubble scales;
\item inflation lasts more than the canonical minimal $\sim 60$ e-folds to explain the features of our observed universe. In this way, in the last 60 e-folds or so there exists a vacuum expectation value of the higher-spin field
which introduces  preferred directions, breaking isotropy.  Here it is important to notice that fluctuations with comoving wavenumber $k$ exit the Hubble radius  after a single realisation of the first  $(N-N_k)$ e-folds of inflation, where $N$ is the total number of e-folds and $N_k$ is the number of e-folds till the end of inflation when such modes exit. They are affected therefore by the value that the infrared  higher-spin field assumes in that single realisation.
The fact that the higher-spin state gets a vacuum expectation value is relevant to our considerations  since it allows, in analogy with the spin-1 case,   a mixing between scalars and higher-spin fields.

\end{enumerate}
 \begin{figure}[!h]
\centering
\includegraphics[width=0.8\textwidth]{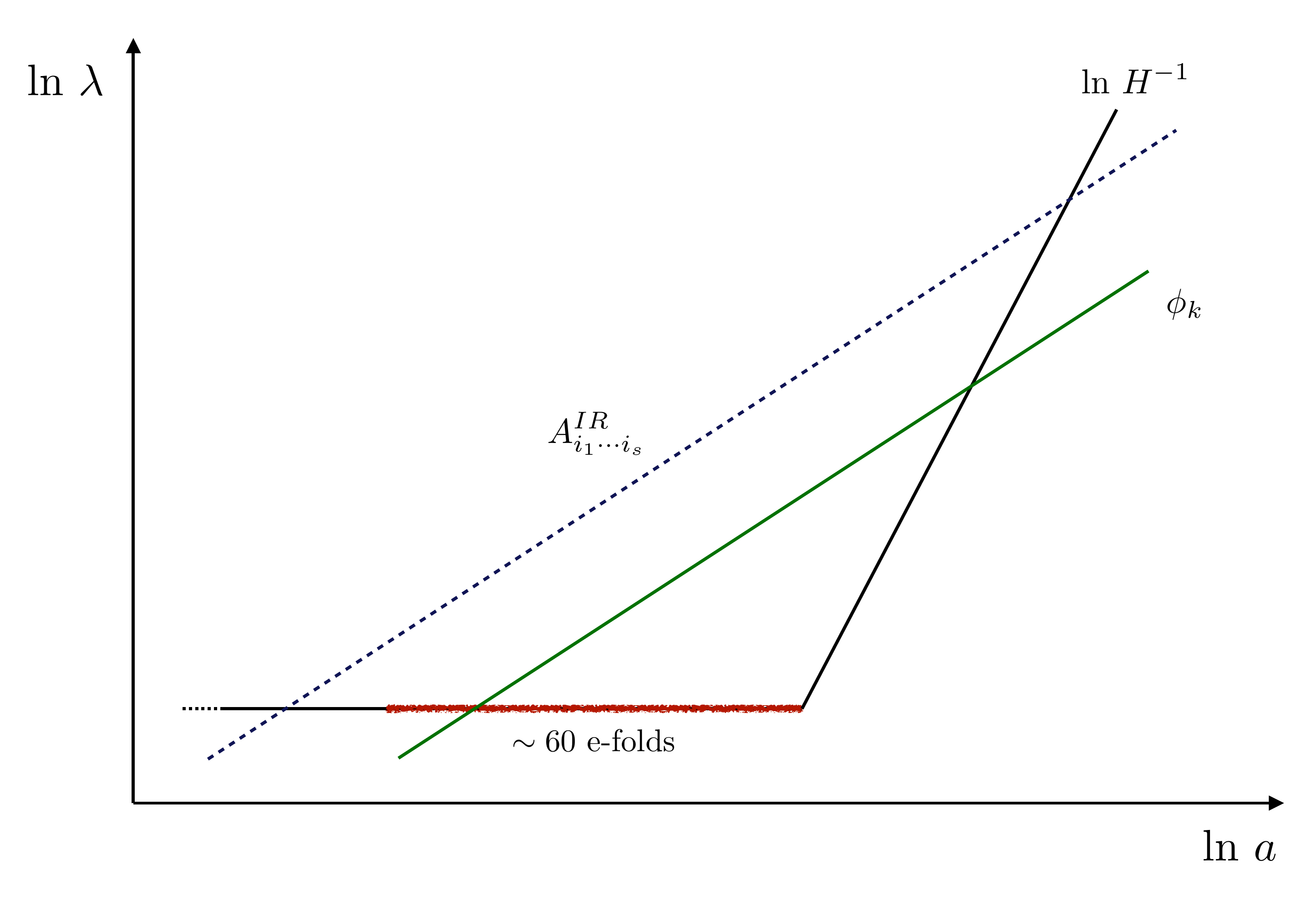}
\caption{Scalar perturbations on cosmologically relevant scales leave the Hubble radius in the presence of a nonvanishing higher-spin infrared background.}\label{img1}
\end{figure}
The goal of this paper  is to characterise the statistically anisotropic signals of the effectively massless higher-spin fields onto the power spectrum, the bispectrum and the trispectrum of the scalar perturbations from the point of view of the dS/CFT$_3$ correspondence when the higher-spin states acquire a vacuum expectation value. 

We will follow  and generalise the techniques developed by  Cardy \cite{cardy}  to compute the   anisotropic corrections to correlation functions in conformal systems  where some components of the energy momentum tensor acquire  non-zero vacuum expectation values.

 A technical tool which will be useful in obtaining our findings will be the Operator Product Expansion (OPE), see for example Refs. \cite{c2,c3}. We will do so by both analysing the case of the partially massless 
spinning states and  extending the findings of Ref. \cite{vector} to the case of fields with spin larger than unity. Our  technique will also allow us  to easily reproduce the  well-known 
results of the spin-1 case \cite{peloso}. 

Since in the case of single-field models of inflation, the inflaton background is still invariant under a  dilations plus a shift of the inflaton field, but not under special conformal transformations, rigorously our considerations will be valid only in the case in which scalar perturbations arise from multi-field models of inflation, e.g. through the curvaton mechanism, where the perturbations are induced by a spectator field $\phi(\tau,\vx)$ during inflation. In such a case gravity is decoupled and  SO(1,4) is a non-linearly realised symmetry of the action in de Sitter. 
In the single-field models
of inflation the scalar fluctuations are sensitive to departures from the special conformal symmetry and we expect that a systematic   breaking of such symmetries could lead  to further constraints at leading order in the slow roll parameters\footnote{The case of a single-field inflation and the role of partially massless higher-spin fields  is discussed in Ref. \cite{am1}. We thank D. Baumann, G. Goon, H. Lee and G.L.  Pimentel for 
sharing their draft with us.}.

%

The paper is organised as follows. In section 2 we study the statistically anisotropic contributions to the scalar two-point correlator from spinning particles, devoting section 3 to the study of the three- and four-point correlators. Section 4 is devoted to the 
analysis restricted to the case of partially massless higher-spin states. Section 5 contains our conclusions.

\section{The statistically anisotropic two-point correlator from spinning particles}
In this section we wish to characterize the anisotropic contributions to the power spectrum of scalar fluctuations induced by the presence of constant super-Hubble modes of higher-spin fields. 

\subsection{The spin-1 case}
Let us start with the known case of a spin-1 field \cite{peloso,vector} which, thanks to a suitable coupling of the form $I(\phi_0)F_{\mu\nu}^2$, has vanishing scaling dimension. 
 
 From now on we call $\phi(\tau,\vx)$ the (spectator scalar)   field which scales towards the boundary $\tau=0$ as $\phi(\tau,\vx)=(-\tau)^{\Delta_\phi}\phi(\vx)$. In this sense, we identify with $\phi(\vx)$ the  scalar primary field 
with conformal weight $\Delta_\phi\simeq 0$. We also assume that the corresponding conformal weight of the spin-1 field on the boundary is close to zero.
The two-point   function $\Big< \phi(\vx_1)\phi(\vx_2)\Big>$ can be worked out in the following way.  Let us perform an inversion  with respect to the origin far from the points $\vx_1$ and $\vx_2$
\begin{eqnarray}
x'^{i}_1= \frac{x_{1}^i}{x_{1}^2}, ~~~x'^{i }_2= \frac{x_{2}^i}{x_{2}^2}, 
\label{transf}
\end{eqnarray} 
such that the new points $x'_1$ and $x'_2$ are brought close to each other, 
\begin{eqnarray}
\Big< \phi(\vx_1)\phi(\vx_2)\Big>= x'^{2\Delta_\phi}_1 x'^{2\Delta_\phi}_2\Big< \phi(\vx_1')\phi(\vx_2')\Big>.
\end{eqnarray}
Since now the points $x'_1$ and $x'_2$ are close, we can perform the OPE expansion on the two-point correlator on the right-hand side. We assume that the
 vector field is $A^i(\vx)$ on the boundary and that its two-point function is  fixed by the symmetries, up to the an overall normalization. More importantly, we also assume that the spin-1 fields acquire a vacuum expectation value since during inflation the infrared long wavelength
 perturbations of the vector field accumulate and provide a classical background for a local observer. 
 
The field $\phi(\vx)$ mixes with the scaling invariant operator 
$A_i(\vx)$  so that in the OPE expansion we expect terms 
of the form
\begin{eqnarray}
\label{p1}
 \phi(\vx_1')\phi(\vx_2')\simeq \frac{c_\phi}{{x'}_{12}^{2\Delta_\phi}}+c_i(x'_{12})A^i(\vx'_2)+c_{ij}(x'_{12})A^iA^j(\vx'_2)+\cdots, \label{ppp}
\end{eqnarray}
where we have adopted the common notation $x^i_{km}=(x^i_k-x^i_m)$.
Due to the gauge invariance symmetry associated to the gauge field $A_i(\vx)$, it is easy to show that the coefficients of the OPE expansion in
Fourier space must satisfy the following relations\footnote{Of course one can start working directly with gauge-invariant fields, in the case at hand with the electric or the magnetic fields \cite{peloso,vector}.}
\be
k^i c_i(k)=k^i c_{ij}(k)=0,
\ee
from which we deduce the relations  $c_i(k)=0$ and  $c_{ij}(k)=c(k)\Pi_{ij}(k)$, where
\be
\label{pij}
\Pi_{ij}(k)=\delta_{ij}-\hat{k}_i\hat{k}_j, \,\,\,\,\,\, \hat{k}_i=\frac{k_i}{k}.
\ee
Undoing now the  inverse transformation to go back to the original coordinates    and performing the Fourier transform  respect to $\vec{x}_{12}$  we get 
\begin{eqnarray}
\Big< \phi_{\vec{k}}\phi_{-\vec{k}}\Big>'= \frac{c_\phi}{k^{3}}
+c(k)(\delta_{ij}-\hat{k}_i\hat{k}_j) \Big< A^iA^j(0)\Big>+\cdots.
\end{eqnarray}
The prime here indicates dropping the Dirac delta for momentum conservation and the factor $(2\pi)^3$. At this stage we pause and provide two comments. First, the OPE, since it is  a short-distance property of the theory, it is not affected by boundary conditions
provided by the vacuum expectation value of the vector field. Secondly, 
the quantity   $\Big< A^iA^j(0)\Big>$ gets two contributions, one from a possible classical value pre-existing   the start of inflation and another inevitable   one generated from the beginning of inflation and caused by the accumulation of the infrared modes. It is seen by a local observer restricted on a finite Hubble volume as a background which breaks isotropy. Following Ref. \cite{peloso} and taking our observed universe as a single realisation of the different ensemble ones, we can write
\be
\Big< A^iA^j(0)\Big>=A_0^2 n^i n^j,
\ee
where $\vec{n}$ identifies the preferred direction of such a single realisation. The typical value of $A_0$  is in such a case the square root of the variance
of the vector field when a given wavenumber $k$ of the scalar fluctuations leave the comoving Hubble radius, that is $A_0\sim H\sqrt{N-N_k}$, where $N$ is the total number of e-folds  of inflation and $N_k$ is the number of e-folds
till the end of inflation when a given comoving wavelength $1/k$ exists the comoving Hubble radius\footnote{Since the associated energy density is of the order of $ H^4 N$, the bound 
$N<(M_{\rm Pl}/H)^2$ must be satisfied to avoid that the energy stored into the infrared modes exceeds the one driving inflation. Notice also that on the last 60 e-folds or so, the classical background evolution in time is suppressed
by powers of $N_k/N\ll 1$ and therefore the background can be taken constant with time with good approximation.}.

In order to fix the coefficient $c(k)$ we use the Ward identity associated to the dilations (we take $\Delta_\phi\simeq 0$)
\begin{eqnarray}
\left(3 +k^i\partial_{k_i}\right)\Big< \phi_{\vec{k}}\phi_{-\vec{k}}\Big>'=0,
\end{eqnarray}
or
\begin{eqnarray}
3c(k)+k\partial_k c(k)=0, 
\end{eqnarray}
which gives 
\begin{eqnarray}
c(k)=\frac{c_1}{k^{3}}. \label{cc}
\end{eqnarray}
Collecting these results we finally obtain the anisotropic contribution to the scalar power spectrum to be
%
%
%
\begin{equation}
\Big<\phi_{\vk}\phi_{-\vk}\Big>'
=\frac{c_\phi}{ k^{3}}\left(1+c_1\sin^2(\hat k\cdot \hat n) +\cdots \right).
\end{equation}
This angle dependence, obtained solely by symmetry arguments,  nicely reproduces the one in  Ref. \cite{peloso}.  

\subsection{The higher-spin case}
One can now
proceed similarly for the fields with spin larger than unity and conformal weight close to zero. Along the same lines, one can write the OPE as 
\begin{eqnarray}
 \phi_{\vec{k}}\phi_{-\vec{k}}\simeq
\frac{c_\phi}{k^{3}}\Big(1+c_{i_1\cdots i_s j_1\cdots j_s}(k) A^{i_1\cdots i_s}A^{j_1\cdots j_s}(0)+\cdots\Big). \label{2pt1}
\end{eqnarray}
Due to an  enhanced symmetry for special values of the parameters in the higher-spin field equations  coupled to suitable functions of time  \cite{noi}, the coefficients $c_{i_1\cdots i_s j_1\cdots j_s}(k)$ satisfy the gauge invariance condition
\begin{eqnarray}
k^{i_{1}} c_{i_1\cdots i_s j_1\cdots j_s}(k)=\cdots=k^{i_{s}} c_{i_1\cdots i_s j_1\cdots j_s}(k)=0, 
\end{eqnarray}
and  are symmetric and traceless in the first and second group of indices as well as in the interchange of the two groups of indices. Clearly, 
 these coefficients are proportional to the  spin helicity sums
 \begin{eqnarray}
c_{i_1\cdots i_s j_1\cdots j_s}(k)=c(k) \sum_\lambda  \epsilon^\lambda_{i_1\cdots i_s}(k) \epsilon^{*\lambda}_{j_1\cdots j_s}(k),
\end{eqnarray}
where  $\lambda$ are the helicities and the polarisation tensors $\epsilon^\lambda_{i_1\cdots i_s}$
are symmetric traceless and satisfy  the relations
\begin{eqnarray}
k^{i_1}\epsilon^\lambda_{i_1\cdots i_s}=0, ~~~~{\epsilon^{\lambda i_1}}_{i_1i_3\cdots i_s}=0.
\end{eqnarray}
\begin{framed}
{\footnotesize
\noindent 
Polarisation tensors of higher-spin fields can  be obtained generalising the notion of polarisation vectors introducing positive and negative energy wave functions
\begin{eqnarray}
\epsilon_\lambda^{i_1\cdots i_s}(k)&=&\sum_{\lambda_1,\cdots,\lambda_s=\pm 1}\delta_{\lambda_1+\cdots+\lambda_s,\lambda}
\sqrt{\frac{2^s(s+\lambda)!(s-\lambda)!}{(2s)!\prod_{i=1}^s(1+\lambda_i)!(1-\lambda_i)!}}\prod_{j=1}^s\epsilon_{\lambda_j}^{i_j}(k),\nonumber\\
\epsilon_\lambda^{*i_1\cdots i_s}(k)&=&\sum_{\lambda_1,\cdots,\lambda_s=\pm 1}\delta_{\lambda_1+\cdots+\lambda_s,\lambda}
\sqrt{\frac{2^s(s+\lambda)!(s-\lambda)!}{(2s)!\prod_{i=1}^s(1+\lambda_i)!(1-\lambda_i)!}}\prod_{j=1}^s\epsilon_{\lambda_j}^{*i_j}(k),
\end{eqnarray}
where $\epsilon_{\lambda}^{i}$ and $\epsilon_{\lambda}^{*i}$ are positive and negative energy wave functions for a spin-1 field,
with
\be
\epsilon_{\lambda}^{*i}=(-1)^\lambda\epsilon_{-\lambda}^{i}.
\ee
It is useful to define the  projector tensor in $d$ dimensions as
\be
\Pi^{ i_1 \cdots i_s j_1 \cdots j_s} (k)
\equiv \sum_{\lambda} \epsilon_\lambda ^{i_1 \cdots i_s} (k) \epsilon^{* j_1 \cdots j_s}_{\lambda } (k). 
\ee
It can be explicitly constructed by using the spin-1 projector tensor $\Pi^{ij}$
\begin{equation}\label{bc}
		\begin{aligned}
		\Pi^{i_1 \cdots i_s j_1 \cdots j_s} (k)=
		 \left( \frac{1}{s!}\right) \sum_{P{(i)}P{(j)}} \left[
		 \sum_{r=0  }^{r\leq\frac{s}{2}} 
C(s,r) \Pi^{i_1 i_2}\Pi^{j_1 j_2} \cdots \Pi^{i_{2r-1} i_{2r}}\Pi^{j_{2r-1} j_{2r}} \prod_{n=2r+1}^{s} \Pi^{i_n j_n} \right],
	\end{aligned}
\end{equation}
where $P(i)P(j)$ stands for independent permutations of $i$ and $j$ sets of indices and, by defining a function $A(m,n)$ so that
\be
\begin{aligned}
&A(m,n) = {{m}\choose{2}} {{m-2}\choose{2}} \cdots {{m-2(n-1)}\choose{2}},
\\
&A(m,n) =0 \ \ \text{for}\ n<0,\ \ \ A(m,n) =1 \ \ \text{for}\ n=0,\ \ \ A(m,n) =0 \ \ \text{for}\ m<2n.
 \end{aligned} 
 \ee
Thus, the coefficients in Eq. (\ref{bc}) are
 \be
 \begin{aligned}
 &C(s,r) =- \left\{\frac{ C(s,r-1)A(s,r-1)A(s-2,r-1) [s-2(r-1)]!}{A(s,r) (s-2r)!\left[A(s,r)-A(s-2,r)+  (d-2) A(s-2,r-1)\right]} \right\},\ \ \ C(s,0)=1.
\end{aligned}
\ee
For instance, for spin-2 in three-dimensions we obtain (we of course sum only over the maximally transverse modes as lower helicity states decay on super-Hubble scales)
\be
\Pi_{i_1i_2}^{j_1j_2} =\frac{1}{2}  \left(\Pi_{i_1}^{j_1}  \Pi_{i_2}^{j_2} + \Pi_{i_1}^{j_2}  \Pi_{i_2}^{j_1}\right) - \frac{1}{2}\Pi_{i_1 i_2}  \Pi^{j_1 j_2}.
\ee
} 
\end{framed}
 \noindent
The coefficient of proportionality can be fixed as before using the Ward identity associated to the dilation symmetry.
Thus, we have
\begin{eqnarray}
\Big< \phi_{\vec{k}}\phi_{-\vec{k}}\Big>' =
\frac{c_\phi}{k^{3}}\Big(1+c_{i_1\cdots i_s j_1\cdots j_s}(k)\Big< A^{i_1\cdots i_s}A^{j_1\cdots j_s}(0)\Big> +\cdots\Big). \label{2pt2}
\end{eqnarray}
We assume  again that there is a background value for the higher-spin field which  defines  a set of
$s$ vectors $n^i_{m}$  with $m=1,\cdots, s$ (not necessarily normalised to unity), such that  
%
\begin{eqnarray}
\Big< A^{i_1\cdots i_s}\Big>= \left[n_1^{(i_1}\cdots n_s^{i_s)}+\cdots\right],
\end{eqnarray}
where the $\cdots$ denote extra terms constructed with Kronecker deltas in such a way to preserve the traceless constraint. In any case, these extra terms are irrelevant when contracted with the polarisation tensors.
We then obtain 
\begin{eqnarray}
\label{frame}
\Big< A^{i_1\cdots i_s}A^{j_1\cdots j_s}(0)\Big>= \left[n_1^{(i_1}\cdots n_s^{i_s)}n_1^{(j_1}\cdots n_s^{j_s)}+\cdots\right],
\end{eqnarray}
The two-point correlator of the bulk field $\phi$ turns out to be 
\begin{equation}
\Big<\phi_{\vk}\phi_{-\vk}\Big>'
= \frac{c_\phi}{  k^{3}}\left(1+c_s\prod_{i=1}^s \sin^{2}(\hat k\cdot \hat{n}_i) +\cdots \right).
\end{equation}
Notice that, in the special case in which all the $\vec{n}_m$ are aligned along a common direction $\vec n$ and a subgroup  SO(2) rotation symmetry is present, the background value for the higher-spin field is
\begin{equation}
	\Big< A^{i_1\cdots i_s}\Big>=A_0 \left[n^{i_1}\cdots n^{i_s}-\frac{1}{2s-1} \left(\delta ^{i_1 i_2} n^{i_3} \cdots n^{i_s} + \text{symm.}\right)+ \cdots\right]
\end{equation}
and the previous equation reduces to the one adopted in Ref. \cite{noianis}
\begin{equation}
\Big<\phi_{\vk}\phi_{-\vk}\Big>'
= \frac{c_\phi}{  k^{3}}\left(1+c_s\sin^{2s}(\hat k\cdot \hat{n}) +\cdots \right).
\end{equation}
From now on, to simplify the expressions, we will be assuming that the vectors $\vec{n}_m$ are aligned along a common direction $\vec n$. 

\section{The statistically anisotropic three- and four-point correlator from spinning particles}
In this section we wish to characterise the anisotropic contributions to the three- and four-point correlators of scalar and tensor fluctuations induced by the presence of constant super-Hubble modes of higher-spin fields. 

\subsection{The spin-1 case}
Let us again start with the known case of the spin-1 field \cite{peloso,vector}. 
The three-point   function $\Big< \phi(\vx_1)\phi(\vx_2) \phi(\vx_3)\Big>$ can be worked out by performing an inversion  with respect to a point $\vz$ 
\cite{cardy}, see Fig. 2
\begin{eqnarray}
x'^{i}= \frac{x^i-z^i}{|\vx-\vz|^2}.
\label{transfz}
\end{eqnarray}
If we take $\vz$ in the vicinity of  $\vx_3$, than  
the new points $x'_1$ and $x'_2$ are brought close to each other and far from $x'_3$. 
We therefore get
\begin{eqnarray}
\Big< \phi(\vx_1)\phi(\vx_2)\phi(\vx_3)\Big>= x'^{2\Delta_\phi}_1 x'^{2\Delta_\phi}_2 x'^{2\Delta_\phi}_3\Big< \phi(\vx_1')\phi(\vx_2')\phi(\vx_3')\Big>.
\end{eqnarray}
This time we find it convenient to perform the OPE by expanding in  powers of the scalar  quantity ${\cal A}(\vx)=A_0^i A_i(\vx)$ 
\begin{figure}[!h]
\centering
\includegraphics[width=0.6\textwidth]{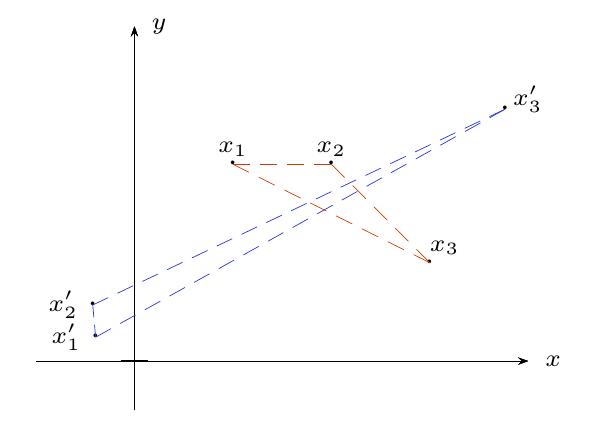}
\caption{The inversion operation around a point $z$ close to $\vx_3$.}\label{img2}
\end{figure}
\begin{eqnarray}
\label{p11}
 \phi(\vx_1')\phi(\vx_2')\simeq\frac{c_\phi}{{x'}_{12}^{2\Delta_\phi}}\Big(1+c_1 {\cal A}(\vx'_2)+c_2 {\cal A}^2(\vx'_2) +\cdots\Big). 
\end{eqnarray}
From this expression we can deduce
\begin{eqnarray}
\label{pp}
\Big< \phi(\vx_1')\phi(\vx_2')\phi(\vx_3')\Big> \simeq \frac{c_2}{{x'}_{12}^{2\Delta_\phi}}\Big< {\cal A}^2(\vx'_2)\phi(\vx_3')\Big>+ \cdots\simeq  \frac{c_2}{{x'}_{12}^{2\Delta_\phi}}\Big< {\cal A}(\vx'_1){\cal A}(\vx'_2)\phi(\vx_3')\Big>+ \cdots. 
\end{eqnarray}
Undoing the inversion operation and going to momentum space, we get the following expression
\begin{eqnarray}
\Big< \phi_{\vec{k}_1}\phi_{\vec{k}_2}\phi_{\vec{k}_3}\Big>'
=c_2 A_0^i A_0^j\Big<A^i_{\vec{k}_1}A^j_{\vec{k}_2}\phi_{\vec{k_3}}\Big>'.
\end{eqnarray}
The three-point correlator on the right-hand side of the last expression is fixed by the Ward identities as found in Ref. \cite{vector} and reads
\begin{eqnarray}
\Big<A^i_{\vec{k}_1}A^j_{\vec{k}_2} \phi_{\vec{k}_3}\Big>'=
\frac{c_{3}}{k_1^{3} k_2^{3}}\Pi^{im}(k_1)\Pi^{mj}(k_2)=
\frac{c_{3}}{k_1^{3} k_2^{3}}
\left(\delta^{ij}-\hat{k}_1^i\hat{k}_1^j-\hat{k}_2^i\hat{k}_2^j+
\hat{k}_1\cdot \hat{k}_2 \, \hat{k}_1^i\hat{k}_2^j\right). 
\end{eqnarray}
 We finally find 
 \begin{eqnarray}
 \label{lll}
\Big< \phi_{\vec{k}_1}\phi_{\vec{k}_2}\phi_{\vec{k}_3}\Big>'
= \frac{c_{\phi\phi\phi}}{k_1^{3} k_2^{3}} 
 \left(1-\cos^2(\hat{k}_1\cdot {\hat{n}})-\cos^2(\hat{k}_2\cdot {\hat{n}})
 +\cos(\hat{k}_1\cdot \hat{k}_2)\cos(\hat{k}_1\cdot {\hat{n}})\cos(\hat{k}_2\cdot {\hat{n}})\right) +{\rm cyclic}.\nonumber\\
 &&
 \end{eqnarray}
This expression again reproduces nicely the findings of Ref. \cite{peloso}. 

\subsection{The higher-spin case}
The calculation of the three-point correlator from higher-spin states goes along the same lines of the previous section and we do not report here in full details. We just notice that 
one  encounters permutations of the spin polarisation sum
\begin{eqnarray}
I_s(\vec{n},\vec{k}_1,\vec{k}_2)=\Big< A^{i_1\cdots i_s}
  A^{j_1\cdots j_s}\Big>\sum_\lambda \Big<  \epsilon^\lambda_{i_1\cdots i_s}(\vec{k}_1) \epsilon^{*\lambda}_{j_1\cdots j_s}(\vec{k}_2)\epsilon^\lambda_{\ell_1\cdots \ell_s}(\vec{k}_1) \epsilon^{*\lambda}_{\ell_1\cdots \ell_s}(\vec{k}_2)\Big>.
\end{eqnarray}
Since the polarisation tensor is traceless we obtain
\begin{eqnarray}
I_s(\vec{n},\vec{k}_1,\vec{k}_2)=A_0^2\,n^{i_1}\cdots n^{i_s}n^{j_1}\cdots n^{j_s}
\sum_\lambda  \epsilon^\lambda_{i_1\cdots i_s}(\vec{k}_1) \epsilon^{*\lambda}_{j_1\cdots j_s}(\vec{k}_2)
\epsilon^\lambda_{\ell_1\cdots \ell_s}(\vec{k}_1) \epsilon^{*\lambda}_{\ell_1\cdots \ell_s}(\vec{k}_2).
\label{is}
\end{eqnarray}
To simplify this expression we 
 average over the directions $n^i$ (an operation which is anyway done when comparing to the observations)
\begin{eqnarray}
I_{s,{\rm av}}(\vec{k}_1,\vec{k}_2) =\int {\rm{d}}\Omega \, I_s(\vec{n},\vec{k}_1,\vec{k}_2).
\end{eqnarray}
Using Eq. (\ref{is}) we find that 
\begin{eqnarray}
I_{s,{\rm av}}(\vec{k}_1,\vec{k}_2) &=&\int {\rm{d}}\Omega\, A_0^2\, n^{i_1}\cdots n^{i_{2s}}
\sum_\lambda  \epsilon^\lambda_{i_1\cdots i_s}(\vec{k}_1) \epsilon^{*\lambda}_{i_{s+1}\cdots i_{2s}}(\vec{k}_2)
\epsilon^\lambda_{\ell_1\cdots \ell_s}(\vec{k}_1) \epsilon^{*\lambda}_{\ell_1\cdots \ell_s}(\vec{k}_2)
\nonumber \\
&=&A_0^2 \sum_\lambda  \epsilon^\lambda_{i_1\cdots i_s}(\vec{k}_1) \epsilon^{*\lambda}_{i_{s+1}\cdots i_{2s}}(\vec{k}_2)
\epsilon^\lambda_{\ell_1\cdots \ell_s}(\vec{k}_1) \epsilon^{*\lambda}_{\ell_1\cdots \ell_s}(\vec{k}_2)
\int {\rm{d}}\Omega\, n^{i_1}\cdots n^{i_{2s}} .
\end{eqnarray}
Using the normalization
\begin{eqnarray}
\int {\rm d}\Omega\, n^in^j=\frac{4}{3}\delta^{ij},
\end{eqnarray}
we find, after some combinatorics, 
\begin{eqnarray}
I_{s,{\rm av}}(\vec{k}_1,\vec{k}_2)=4 A_0^2\frac{s!}{(2s+1)!!}
\sum_\lambda  \Big(\epsilon^\lambda_{i_1\cdots i_s}(\vec{k}_1) \epsilon^{*\lambda}_{i_{1}\cdots i_{s}}(\vec{k}_2)\Big)^2.
\end{eqnarray}
To calculate, for instance,  the sum over helicities for $s=1$, we may use a coordinate system where the polarisation $\epsilon^+(\vec{k}_1)$ and $\epsilon^+(\vec{k}_2)$ are aligned. In this system, it is easy to find that 
\begin{eqnarray}
\sum_\lambda  \Big(\epsilon^\lambda_{i}(\vec{k}_1) \epsilon^{*\lambda}_{i}(\vec{k}_2)\Big)^2=
1+\cos^2\left(\hat{k}_1\cdot \hat{k}_2\right),
\end{eqnarray}
which of course reproduces the direction average of the angle dependence in the expression (\ref{lll}).
For a generic spin $s$ we find
\begin{eqnarray}
\sum_\lambda  \Big(\epsilon^\lambda_{i_1\cdots i_s}(\vec{k}_1) \epsilon^{*\lambda}_{i_{1}\cdots i_{s}}(\vec{k}_2)\Big)^2=1+\cos^{2s}\left(\hat{k}_1\cdot \hat{k}_2\right),
\end{eqnarray}
and therefore
\begin{eqnarray}
I_{s,{\rm av}}(\vec{k}_1,\vec{k}_2)=4 A_0^2\frac{s!}{(2s+1)!!}\left(1+\cos^{2s}\left(\hat{k}_1\cdot \hat{k}_2\right)\right).
\end{eqnarray}
The averaged three-point correlator in the presence of higher-spin fields  will be therefore of the form
\begin{eqnarray}
 \label{llll}
\Big< \phi_{\vec{k}_1}\phi_{\vec{k}_2}\phi_{\vec{k}_3}\Big>'_{\rm av}
= \frac{c_{\phi\phi\phi}}{k_1^{3} k_2^{3}} 
 \left(1+\cos^{2s}\left(\hat{k}_1\cdot \hat{k}_2\right)\right) +{\rm cyclic}.
 \end{eqnarray}
\subsection{The three-point correlator involving higher-spin fields}
We want to use the OPE to write the three-point function $\langle A_{\vec{k}_1}^{ i_1 \cdots i_s} \phi_{\vec{k}_2}\phi_{\vec{k}_3} \rangle$. 
Using the inversion in the coordinates space, the OPE for a couple of scalar operators will have the form:
\begin{equation}
	\label{OPE1}
	\phi (\vx_2') \phi (\vx_3') \simeq \frac{1}{x_{23}^{\prime 2 \Delta _\phi}} \left( c_0 + c_1 \mathcal{A} ( \vec{ x}_{3}' ) + \cdots \right) \simeq \frac{1}{x_{23}^{\prime  2 \Delta _\phi}} \left( c_0 + c_1 A_{0}^{j_1 \cdots j_s} A_{j_1 \cdots j_s} (\vec {x}_3 ')+\cdots \right).
\end{equation}
Thus, we can write
\be
\Big\langle A^{i_1 \cdots i_s} (\vx_1')\phi (\vx_2') \phi (\vx_3') \Big\rangle = \frac{c_1}{x_{23}^{\prime  2 \Delta _\phi}}  A_{0}^{j_1 \cdots j_s}  \Big\langle  A^{i_1 \cdots i_s} (\vx_1') A_{j_1 \cdots j_s} (\vx_3') \Big \rangle,
\ee
which, performing the inversion backwards and transforming to the momentum space, becomes
\be\label{step}
\Big\langle A^{i_1 \cdots i_s} _{\vec{k}_1} \phi _{\vec{k}_2}\phi _{\vec{k}_3} \Big\rangle' &=&
\frac{c_2}{k_2^{3}}A_{0}^{j_1 \cdots j_s}   \Big\langle  A^{i_1 \cdots i_s} (\vk_1) A_{j_1 \cdots j_s} (-\vk_1) \Big \rangle
\nonumber \\
&=& \frac{c_{3}}{k_1^3 k_2^{3}}  A_{0}^{j_1 \cdots j_s}     \sum_{\lambda} \epsilon_\lambda ^{i_1 \cdots i_s} (\vec{k}_1) \epsilon^*_{\lambda j_1 \cdots j_s} (\vec{k}_1) + \text{cyclic} .
\ee
In the last step leading to \eqref{step} we have used the Ward identity associated to dilation isometry and the fact that the tensorial structure of the two higher-spin two-point function is given by the projector tensor $\Pi^{ i_1 \cdots i_s}_{ j_1 \cdots j_s} ({k})$.
Now we assume that  the background vacuum expectation value of the higher-spin field  has the usual structure in terms of one unit vector $\vec n$
and we get
\be\label{Apipi}
\Big\langle A^{i_1 \cdots i_s} _{\vec{k}_1} \phi _{\vec{k}_2}\phi _{\vec{k}_3} \Big\rangle' = \frac{c_{A\phi\phi}}{k_1^3 k_2^{3}}  \left(n^{j_1} \cdots n^{j_s}  \Pi^{ i_1 \cdots i_s}_{ j_1 \cdots j_s} ({k}_1) \right)+ \text{cyclic}.
\ee
In the simplified case in which we consider $s=2$, we can expand the tensorial structure of \eqref{Apipi}
	\be
	\begin{aligned}
		n^in^j \Pi_{ij}^{kl}({k_1}) =n^in^j \left[\frac{1}{2}  \left( \Pi_i^k  \Pi_j^l+ \Pi_i^l  \Pi_j^k\right) -\frac{1}{2} \Pi_{ij}  \Pi^{kl}\right]=
		\\
		=v^k(\hat {k}_1, \hat n) v^l(\hat {k}_1, \hat n) - \frac{1}{2} \sin ^2 ( \hat {k}_1 \cdot \hat n ) \Pi^{kl} ({k}_1),
	\end{aligned}
		\ee
		where 
		\be
		v^k(\hat {k}, \hat n) = \hat {n}^k - \hat {k} ^ k \cos( \hat k \cdot \hat n).
		\ee
\subsection{The statistically anisotropic three-point correlator tensor-scalar-scalar from spinning particles}
Using the OPE we can also work out the statistically anisotropic contribution to the three-point correlator $\langle\gamma\phi\phi\rangle$ involving the massless spin-2 graviton state. We start from Eq. (\ref{pp}) to get 
\begin{eqnarray}
\label{pp1}
\Big< \phi(\vx_1')\phi(\vx_2')\gamma^{ij}(\vx_3')\Big> \simeq \frac{c_2}{{x'}_{12}^{2\Delta_\phi}}\Big< {\cal A}^2(\vx'_2)\gamma^{ij}(\vx_3')\Big>+ \cdots\simeq  \frac{c_2}{{x'}_{12}^{2\Delta_\phi}}\Big< {\cal A}(\vx'_1){\cal A}(\vx'_2)\gamma^{ij}(\vx_3')\Big>+ \cdots. 
\end{eqnarray}
Undoing the inversion operation and going to momentum space, we obtain
\begin{eqnarray}
\label{stanc}
\Big< \phi_{\vec{k}_1}\phi_{\vec{k}_2}\gamma^{ij}_{\vec{k}_3}\Big>'
=c_2 A_0^{i_1\cdots i_s} A_0^{j_1\cdots j_s}\Big<A^{i_1\cdots i_s}_{\vec{k}_1}A^{j_1\cdots j_s}_{\vec{k}_2}\gamma{ij}_{\vec{k_3}}\Big>'.
\end{eqnarray}
For instance, for the spin-1 case, using the same techniques of Ref. \cite{peloso}  we get that the statistical anisotropic contribution reads
\be
\Big<\phi_{\vec{k}_1}\phi_{\vec{k}_2}\gamma^\lambda_{\vec{k_3}}\Big>'\supset\frac{c_{\phi\phi\gamma}}{k_2^3 k_3^3}\epsilon_{ij}^\lambda(\hat{k}_3)v^i(\hat{k}_1,\hat n)v^j(\hat{k}_2,\hat n).
\ee
For higher-spin fields, this expression generalizes to 
\be
\Big<\phi_{\vec{k}_1}\phi_{\vec{k}_2}\gamma^\lambda_{\vec{k_3}}\Big>'\supset\frac{c_{\phi\phi\gamma}}{k_2^3 k_3^3}\epsilon_{ij}^\lambda(\hat{k}_3)v^{i i_2\cdots i_s}(\hat{k}_1,\hat n)v^{j}_{ i_2\cdots i_s}(\hat{k}_2,\hat n),
\ee
where
\be
v^{i_1\cdots i_s}(\hat{k},\hat n)=n^{j_1}\cdots n^{j_s}\Pi^{i_1\cdots i_s}_{j_1\cdots j_s}(\hat k).
\ee

\subsection{The statistically anisotropic four-point correlator from spinning particles in the collapsed limit}
We may also consider the four-point scalar correlator in the collapsed limit, that is the configuration in real space where two pairs of points, say
$\vx_1$, $\vx_2$ and $\vx_3$, $\vx_4$ are very far from each other.  Let us therefore consider the OPE expansion (\ref{OPE1}) as well as the one for the other (34) channel at the coincident point
to get 
\be
\Big<\phi(\vx_1)\phi(\vx_2)\phi(\vx_3)\phi(\vx_4)\Big>\simeq\frac{1}{x_{12}^{2\Delta_\phi}x_{34}^{2\Delta_\phi}}\left(c_0+c_1A_{0}^{i_1 \cdots i_s} A_{0}^{j_1 \cdots j_s}\Big< A_{i_1 \cdots i_s} (\vec {x}_1) A_{j_1 \cdots j_s} (\vec {x}_3)\Big>\right)
+\cdots. \label{ab}
\ee
The statistically anisotropic contribution to the four-point correlator therefore reads in momentum space with  $\Delta_\phi\simeq 0$
\begin{eqnarray}
\Big< \phi_{\vk_1}\phi_{\vk_2}\phi_{\vk_3}\phi_{\vk_4}\Big>^\prime \supset \frac{A_{0}^{i_1 \cdots i_s} A_{0}^{j_1 \cdots j_s}}{ k_2^{3} k_4^{3}}
\Big< A_{i_1 \cdots i_s}(\vec {k}_{12}) A_{j_1 \cdots j_s}(-\vec {k}_{12})\Big>
+
\,\,{\rm cyclic},\,\,\,~~(\vk_{12}=\vk_1\!+\!\vk_2\simeq \vec{0}).
\end{eqnarray}
Assuming again the background for the higher-spin to be determined by a single vector 
$\vec n$, , we find then 
 \be
 \label{asd}
\Big< \phi_{\vk_1}\phi_{\vk_2}\phi_{\vk_3}\phi_{\vk_4}\Big>^\prime \supset \frac{c_{\phi\phi\phi\phi}\sin^{2s}(\hat k_{12}\cdot \hat n)}{k_{12}^{3} k_2^{3} k_4^{3}}+
\,\,{\rm cyclic},\,\,\,\,\,\,\,\,\,\,(\vk_{12}=\vk_1+\vk_2\simeq \vec{0}).
\ee

\section{The statistically anisotropic correlators from partial higher-spin  massless states}
Massless particles of spin $s$ in four-dimensional Minkowski space-time possess only  helicities $\pm s$. On the other hand, massive fields have helicities belonging to the set $-s, -s +1, ... , s-1, s$. 
Analogously, in de Sitter space-time  massive  fields are allowed, although there exist additional fields, named ``partially massless" fields \cite{p1,p2,p3,p4,p5,witten} , which do not belong to the aforementioned categories. In particular, fields with mass
\begin{equation}
\label{pn}
	m^2= H^2 \left[ s(s-1) - p(p+1)\right]
\end{equation}
have helicities $-s, -s+1, ..., s-1,s$ where the helicities $-p, -p+1, ... , p-1, p$ have been removed for any $p \leq s-2$. 
Fields of this nature are described by totally symmetric tensors $A_{\mu_1 ... \mu_s}$ whose linear action is invariant under the gauge transformation 
\begin{equation}
\delta A_{\mu_1\cdots\mu_s}=D_{\mu_1}\cdots D_{\mu_{s-p}}\xi_{\mu_{s-p+1}\cdots \mu_s}+\cdots, 
\end{equation}
where $D_\mu$ is the covariant derivative, $\xi_{\mu_{s-p+1}\cdots \mu_s}$ is the gauge parameter and the ellipsis stands for further terms coming from the symmetrization of indices and contributions of terms with a number of derivatives fewer than $s-p-1$.

It is straightforward to notice that spin-1 fields in de Sitter do not possess partially massless states. The usual vector gauge invariance 
\begin{equation}\label{gauge-spin1}
	\delta A_\mu = D_\mu \xi
\end{equation}
can be acquired only be setting $m^2 =0$. The first non-trivial case is the spin-2 field for which partial massless stases having four degrees of freedom exists with $p=0$ and $m^2 = 2 H^2$. The associated gauge invariance can be written explicitly as \cite{witten}
\begin{equation}
\label{vb}
	\delta A_{\mu\nu} = D_\mu D_\nu \xi + H^2 g_{\mu\nu} \xi.
\end{equation}
Using the dS/CFT$_3$ correspondence, as a massless spin-$s$ field has its correspondent rank-$s$ conserved symmetric tensor in the boundary theory \cite{witten}, the  partially massless fields $A_{\mu_1 ... \mu_s}$ correspond to partially conserved currents $L^{\mu_1 ... \mu_s}$ on the boundary. In other words, since the coupling 
\begin{equation}
	\int_{\tau=0} L^{\mu_1 \cdots \mu_s}A_{\mu_1 \cdots \mu_s}
\end{equation}
is gauge invariant, by applying the transformation to the higher-spin field and through straightforward manipulations,  the following condition must be required on the currents $L^{\mu_1 ... \mu_s}$
\begin{equation}\label{gauge-cond-L}
	D_{\mu_1}\cdots D_{\mu_{s-p}}L^{\mu_1 \cdots \mu_s} =0.
\end{equation}
Terms of lower order in derivatives which are proportional to the curvature of the boundary  have been neglected. Using the fact that the boundary at $\tau \rightarrow 0$ is flat,  one can interpret covariant derivatives in  Eq. \eqref{gauge-cond-L} as ordinary derivatives
\begin{equation}
\partial_{i_1} \cdots \partial_{i_{s-p}} L^{i_1 \cdots i_s}=0. 
\end{equation}
This condition can be used to characterize better the partial massless states. Notice that the dual field $L^{i_1 \cdots i_s}$ is sourced by the boundary value of the corresponding higher-spin field $A_{i_1 \cdots i_s}(\vx)$ and therefore
the latter satisfies the same gauge transformation 
\begin{equation}
\delta A_{i_1 \cdots i_s}(\vx)=D_{i_1}\cdots D_{i_{s-p}}\xi_{i_{s-p+1}\cdots i_s}+\cdots.
\end{equation}
Let us enter in more detail about the conformal algebra.
It is well  known that the conformal group SO(1,4) has ten generators, namely translations $\mathscr{P}_i$, dilations 
$\mathscr{D}$, special conformal transformations $\mathscr{K}_i$ and space rotations $\mathscr{L}_{ij}$. The corresponding  
 conformal algebra is defined by the following commutation rules
\be
\label{conf}
&&[\mathscr{D},\mathscr{P}_i]=i \mathscr{P}_i \label{DP}, \\
&&[\mathscr{D},\mathscr{K}_i]=-i\mathscr{K}_i \label{DK}, \\
&&[\mathscr{K}_i,\mathscr{P}_j]=2i\Big{(}\delta_{ij}\mathscr{D}-\mathscr{L}_{ij}\Big{)}, \\
&&[\mathscr{L}_{ij},\mathscr{P}_k]=i\Big{(}\delta_{jk}\mathscr{P}_i-\delta_{ik}\mathscr{P}_j\Big{)}, \\
&&[\mathscr{L}_{ij},\mathscr{K}_k]=i\Big{(}\delta_{jk}\mathscr{K}_i-\delta_{ik}\mathscr{K}_j\Big{)}, \\
&&[\mathscr{L}_{ij},\mathscr{D}]=0, \label{LD}\\
&&[\mathscr{L}_{ij},\mathscr{L}_{kl}]=i\Big{(}\delta_{il} \mathscr{L}_{jk}-\delta_{ik}\mathscr{L}_{jl}+\delta_{jk}\mathscr{L}_{il}-\delta_{jl}\mathscr{L}_{ik}\Big{)\label{LL}}.
\ee
Due to the properties of the algebra, it is useful to label the irreducible representations in terms of their scaling dimension $\Delta$ and their spin $s$ taking advantage of the commutativity of the generators $\mathscr{D}$ and $\mathscr{L}_{ij}$ 
\be
	\mathscr{D} |\Delta ,s \rangle = -i \Delta  |\Delta ,s \rangle,\\
	\mathscr{L}_{ij}|\Delta ,s \rangle_{l} = \left( \Sigma_{ij}\right)^{l'}_{l}  |\Delta ,s \rangle_{l'}.
\ee
Moreover, one can think of the operators $\mathscr{P}_i$ and $\mathscr{K}_i$ as being the raising and lowering (ladder) operators with respect to the scaling dimension. Indeed, using the algebra one finds:
\be
\mathscr{D} \mathscr{P}_i | \Delta,s \rangle = i (\Delta+1) \mathscr{P}_i| \Delta,s \rangle,
\\
\mathscr{D} \mathscr{K}_i | \Delta,s \rangle= i (\Delta-1) \mathscr{K}_i| \Delta,s \rangle.
\ee
In what follows we also use the property for which, in radial quantization, $\mathscr{P}_i^\dagger=\mathscr{K}_i$. 
Taking advantage of the operator-state correspondence of  the CFT$_3$ which relates the operator $L^{\mu_1 \cdots \mu_s}$ to a state $|L^{\mu_1 \cdots \mu_s} \rangle$, we can analyse condition \eqref{gauge-cond-L} further \cite{witten}.
For instance, we already argued that for spin-1 condition \eqref{gauge-spin1} for the bulk field becomes $\partial_i L^{i}=0$ for the boundary field. Using the realization of the translation operator $\mathscr{P}_i$, we observe that the first descendant state of $|L^i \rangle$, which is
$\mathscr{P}_i |L^i \rangle$, is required to be a null vector. Explicitly, one finds
\begin{eqnarray}
|| \mathscr{P}_i |L^i \rangle ||^2&=&\Big< L^r \Big|\mathscr{K}_r \mathscr{P}_i \Big| L^i \Big>\nonumber\\
&=&\Big< L^r \Big |[\mathscr{K}_r,\mathscr{P}_i] \Big|L^i \Big>\nonumber\\
&=&(h-2)\Big< L^i|L_i \Big> =0\,,
\end{eqnarray}
which constraints the conformal dimension of $L^i$ to be $h=2$.
Going to the case of generic spins-$s$,  due to the fact that $L^{i_1 \cdots i_s}$ is partially conserved as in Eq. \eqref{gauge-cond-L}, the corresponding descendant must have vanishing norm
\begin{eqnarray}
|| \mathscr{P}_{i_1}\mathscr{P}_{i_2}\cdots \mathscr{P}_{i_m}|{L}^{i_1 i_2\cdots i_s}\rangle||^2&=&\Big< 
L^{j_1 j_2\cdots j_s}\Big|\mathscr{K}_{j_1}\mathscr{K}_{j_2}\cdots \mathscr{K}_{j_s} \mathscr{P}_{i_1}\mathscr{P}_{i_2}\cdots \mathscr{P}_{i_s}\Big|L^{i_1 i_2\cdots i_s}\Big>\nonumber\\
&\sim& m!(h-(s+1))(h-s)\cdots(h-(2+s-m))|| L^{i_1 i_2\cdots i_s}\rangle||^2.\nonumber\\
&&
\label{mji}
\end{eqnarray}
In particular, we want to impose that the $m^{\rm th}$-descendant with $m=r$ to be null, but nonvanishing if $m <r$ in order to match condition \eqref{gauge-cond-L}. This is achieved if 
\beq\label{cond-dim}
h= 2 + s - r.
\eeq
 Therefore, we can conclude that partially massless spin-$s$ fields in four dimensional de Sitter space-time correspond to states on the boundary theory with conformal dimension given by condition \eqref{cond-dim}, where $r$ ranges over the set $2, \cdots, s$.

Now, for  $s>2$ partial massless states always exist  for which $\Delta=0$. To show this we use the generic relation (\ref{aa}) between masses, Hubble rate and scaling dimensions. 
 On the boundary $\tau=0$, the dominant scaling dimension  for  
 the partial massless states becomes
\beq
\D=\frac{3}{2}- \sqrt{\left(s-\frac{1}{2}\right)^2-s(s-1)+p(p+1)}=\frac{3}{2}- \sqrt{\frac{1}{4}+p(p+1)}.
\eeq
For the choice
\beq
p(p+1)=2\Rightarrow p=1\Rightarrow m^2=H^2 \left[s(s-1)-2\right]\,\,\,{\rm for}\,\,\, s>2,
\eeq
one gets $\Delta=0$ and since $p\leq s-2$, such state for which $p=1$ always exists. For instance, 
the case $s=3$ possesses  two  partially massless states, one for $p=1$ and $\Delta=0$ and the other for $p=0$ and $\Delta=1$.
 In Eq. (\ref{mji}) they correspond to $m=s-p=2$ and  $m=s-p=3$, respectively. For  $m=r=2$, one has $h=2+3-2=3$, corresponding  to $\Delta=0$; for $m=r=3$, one has 
$h=2+3-3=2$, corresponding  to $\Delta=1$. These values  reproduce the conformal weights of the partially massless states.
Of course we are interested in those 
partially massless states for which  $\Delta=0$ as their corresponding fluctuations remain constant  on super-Hubble scales. 

\begin{framed}
{\footnotesize
\noindent 
For $s>3$, there exists always  helicity states for which the corresponding $\Delta<0$. It is not clear if this is a problem as one should deal in any case with gauge-invariant quantities. For instance, 
due  to the   gauge transformation, $A_{\mu\nu}$ is not observable and one should construct a gauge-invariant tensor. Such off-shell tensor in de Sitter space is \cite{gi}
\begin{eqnarray}
\label{gi}
\mathscr{F}_{\mu\nu\rho}=D_\mu A_{\nu\rho}-D_\nu  A_{\mu\rho}. \label{F}
\end{eqnarray}
 Therefore gauge-invariant quantities will contain  in principle a sufficient number of derivatives to cancel the bad behaviour of the non gauge-invariant fields.
 }
 \end{framed}

\subsection{The two-,  three- and four-point correlators from partial massless particles}
We work with  the  partially massless states, which provide the representation of de Sitter  isometry group and  coincide with those of CFT$_3$ on super-Hubble scales. 
We assume there is a  mixing 
between the scalar field $\phi$ and the partially massless fields generated  by  a nonvanishing  background  of the higher-spin fields (so that    our  universe is a single realisation of the different ensembles) and by  suitable couplings between the  quadratic and cubic terms in the higher-spin fields and the scalar\footnote{Explicit gauge-invariant cubic terms for partially massless fields have been constructed,  in (A)dS in, for instance,  Refs. \cite{f1,ft,f3}.}. Alternatively, one can imagine to couple the scalar field to a gauge-invariant higher-spin field (a proper generalisation of  $\mathscr{F}_{\mu\nu\rho}$ in Eq. (\ref{gi})) which
slightly away from exact de Sitter will contain the necessary couplings.
On super-Hubble scales the  partial masslessness of depth $n$ is defined by the condition\footnote{The depth $n$ is defined to be $s-p$ where $p$ defines the missing helicities, see Eq. (\ref{pn}).}
\begin{eqnarray}
k^{i_1}\cdots k^{i_n}A_{i_1\cdots i_s}=0, ~~~n\leq s. \label{ss0}
\end{eqnarray}
The condition (\ref{ss0}) projects out the helicities 
$(-s+n,-s+n+1,\cdots,0,\cdots,s-n-1,s-n)$ and therefore only the helicities $(-s,\cdots,-s+n-1,s-n+1,\cdots,s)$ are physical. 

In the end, we are interested in partially massless spin-$s$ fields which posses a conformal dimension $\Delta=0$. The lowest possible spin is in fact $s=3$ with $p=1$. We provide the explicit construction for this particular case and then we generalise the results for partially massless field of spin $s$. 
In this case, the theory enjoys the gauge invariance 
\begin{eqnarray}
A_{ijk}(k)\to A_{ijk}(k)+k_ik_j \xi_k(k)+ \cdots
 \end{eqnarray} 
 since we are only interested in fields with scaling dimension $\Delta =0$.
The contribution to the two-point function can be written as 
\begin{eqnarray}
\phi_{\vec{k}}\phi_{-\vec{k}}\simeq
\frac{c_\phi}{k^{3}}+c^{i_1 i_2 i_3 j_1 j_2 j_3}(k)A_{i_1 i_2 i_3}A_{ j_1 j_2 j_3}(0)+\cdots ,
\end{eqnarray}
where $c^{i_1 i_2 i_3 j_1 j_2 j_3}$ is symmetric-traceless in $i_1 i_2 i_3 $ and $j_1 j_2 j_3$ independently and satisfies
\begin{equation}\label{pp.1}
\begin{aligned}
	&c^{i_1 i_2 i_3 j_1 j_2 j_3}(k)k_{i_1} k_{i_2}k_{j_1} k_{j_2}= 0, 
\end{aligned}	
\end{equation}
due to gauge invariance, as well as being symmetric with respect to the exchange of the two groups of indices.
Using Eq. \eqref{pij},
which implies $k_i \Pi^{ij}=0$, it is easy to verify that (\ref{pp.1}) is solved in terms of $\hat{k}_i$ and the projection tensor $\Pi^{ij}$ by 
\begin{equation}
\begin{aligned}
c^{i_1 i_2 i_3 j_1 j_2 j_3}(k)=&P_1(k) \Big[ \hk^{i_1} \hk^{j_1} \Big(\Pi^{i_2 j_2}\Pi^{i_3 j_3}+\Pi^{i_2 j_3}\Pi^{i_3 j_2}-\Pi^{i_2 i_3}\Pi^{j_2j_3}\Big)+ \text{symm.}\Big] +
\\
&P_2(k)\Big(f^{i_1 i_2 i_3 j_1 j_2 j_3} - \frac{1}{2}g ^{i_1 i_2 i_3 j_1 j_2 j_3}\Big).
\end{aligned}
\end{equation}
where $P_1(k)$ and $P_2(k)$ are two arbitrary functions to be fixed by dilation symmetry and  in order to simplify the notation, we have defined
\begin{equation}
\begin{aligned}
		f^{i_1 i_2 i_3 j_1 j_2 j_3} = \Pi^{i_1 j_1}\Pi^{i_2 j_2}\Pi^{i_3 j_3}  +
		\Pi^{i_1 j_2}\Pi^{i_2 j_1}\Pi^{i_3 j_3} +\Pi^{i_1 j_3}\Pi^{i_2 j_1}\Pi^{i_3 j_2}
		\\
		+\Pi^{i_1 j_1}\Pi^{i_2 j_3}\Pi^{i_3 j_2}+\Pi^{i_1 j_2}\Pi^{i_ 2j_3}\Pi^{i_ 3j_1}  +\Pi^{i_1 j_3}\Pi^{i_2 j_2}\Pi^{i_3 j_1} 
\end{aligned}
\end{equation}
and 
\begin{equation}
\begin{aligned}
	g^{i_1 i_2 i_3 j_1 j_2 j_3}=&
\Pi^{i_1 i_2}\Pi^{j_1 j_2}\Pi^{i_3 j_3}  +\Pi^{i_1 i_3}\Pi^{j_1 j_2}\Pi^{i_2 j_3} + \Pi^{i_2 i_3}\Pi^{j_1 j_2}\Pi^{i_1 j_3} 
		\\
+&\Pi^{i_1 i_2}\Pi^{j_1 j_3}\Pi^{i_3 j_2}  +\Pi^{i_1 i_3}\Pi^{j_1 j_3}\Pi^{i_2 j_2} +\Pi^{i_2 i_3}\Pi^{j_1 j_3}\Pi^{i_1 j_2}
		\\
+&\Pi^{i_1 i_2}\Pi^{j_2 j_3}\Pi^{i_3 j_1} +\Pi^{i_1 i_3}\Pi^{j_2 j_3}\Pi^{i_2 j_1}+\Pi^{i_2 i_3}\Pi^{j_2 j_3}\Pi^{i_ 1 j_1}.
\end{aligned} 
\end{equation}
It is easy to check that it is not possible to respect condition \eqref{pp.1} with a term of the form $ \hat k^{i_1} \hat k^{j_1} \hat k^{i_2} \hat k^{j_2}\Pi^{i_3j_3}$ as well as the term  $\hat k^{i_1} \hat k^{j_1} \hat k^{i_2} \hat k^{j_2}\hat k^{i_3} \hat k^{j_3}$.
Therefore, using
\begin{eqnarray}
\Big< A_{i_1 i_2 i_3}A_{j_1 j_2 j_3}(0)\Big>=A_0^2
\Big[n_{i_1}n_{i_2} n_{i_3} n_{j_1}n_{j_2} n_{j_3} +\cdots\Big],
\end{eqnarray}
we get
\begin{equation}
\Big<\phi_{\vec{k}}\phi_{-\vec{k}}\Big>' = 
\frac{c_{\phi}}{k^{3}}\Big[ 1+p_1 \sin^{6}(\hat{k}\cdot \hat n) + p_2 \sin^{4}(\hat{k}\cdot \hat n)\Big]. 
\end{equation}
Generalizing this result, the  contribution from the partially massless higher-spin field of depth $n$ to the   two-point function of a scalar field is
\begin{eqnarray}
\phi_{\vec{k}}\phi_{-\vec{k}}=
\frac{c_\phi}{k^3}+c^{i_1\cdots i_sj_1\cdots j_s}(k)
A_{i_1\cdots i_s}A_{j_1\cdots j_s}(0)+\cdots, \label{2pts}
\end{eqnarray}
where 
\begin{eqnarray}
c^{i_1\cdots i_sj_1\cdots j_s}(k)&=&
P_1(k)\Big( \hat k^{i_1}\cdots \hat k^{i_{n-1}} \hat k^{j_1}\cdots  \hat k^{j_{n-1}}\Pi^{i_nj_n}\cdots \Pi^{i_sj_s}+\cdots \Big)\nonumber  \\
&+&P_2(k)\Big( \hat k^{i_1}\cdots \hat k^{i_{n-2}} \hat k^{j_1}\cdots \hat k^{j_{n-2}}\Pi^{i_{n-1}j_{n-1}}\cdots \Pi^{i_sj_s}+\cdots \Big)\nonumber \\
&+&\vdots ~~\nonumber \\
&+&P_n(k)\Big( \Pi^{i_1j_1}\cdots \Pi^{i_sj_s}+\cdots \Big).
\end{eqnarray}
The coefficients $P_1(k),\cdots,P_n(k)$ are straightforwardly fixed by the Ward identity from dilations.
We indicate the vacuum expectation values of the partially massless higher fields by Eq. (\ref{frame}).
Thus, we find 
%
\begin{align}
\Big<\phi_{\vec{k}}\phi_{-\vec{k}}\Big>'&=
\frac{c_\phi}{k^{3}}\bigg[1+\tilde p_1\sin^{2s}(\hat{k}\cdot \hat n)+\tilde p_2\sin^{2(s-1)}(\hat{k}\cdot \hat n)\cos^2(\hat{k}\cdot \hat n)+\nonumber \\
&\qquad\qquad\qquad\cdots
+\tilde p_n\sin^{2(s-n+1)}(\hat{k}\cdot \hat n)\cos^{2(n-1)}(\hat{k}\cdot \hat n)\bigg]\nonumber \\
&=
\frac{c_\phi}{k^{3}}\left[1+\sum_{m=0}^{n-1}p_{m+1}\sin^{2(s-m)}(\hat{k}\cdot \hat n)\right].
\end{align}
Finally, the contribution of the partial massless higher-spin fields to the averaged
three-point function is given again by summing over 
the helicities $(-s,\cdots,-s+n+1,s-n-1,\cdots, s)$. The result for the contribution of partially massless higher-spin fields of depth $n$ turns out then to be
\begin{eqnarray}
\Big< \phi_{\vec{k}_1}\phi_{\vec{k}_2}\phi_{\vec{k}_3}\Big>'_{\rm av}& =  &
\frac{c_{\phi\phi\phi}}{k_1^3 k_2^3}
\bigg\{1+\cos^{2s}(\hat{k}_1\cdot \hat{k}_2 )+p_1 \cos(\hat{k}_1\cdot \hat{k}_2)\Big[1+\cos^{2s-1}(\hat{k}_1\cdot \hat{k}_2)\Big]+
\cdots\nonumber \\
&+&p_{s-n+1}\cos^{s-n+1}(\hat{k}_1\cdot \hat{k}_2)\Big[1+\cos^{s+n-1}(\hat{k}_1\cdot \hat{k}_2)\Big]\bigg\}+{\rm cyclic}\nonumber\\
&=&\frac{c_{\phi\phi\phi}}{k_1^3k_2^3}
\sum_{m=n-1}^{s} p_{s-m}\cos^{s-m}(\hat{k}_1\cdot \hat{k}_2)\Big[1+\cos^{s+m}(\hat{k}_1\cdot \hat{k}_2)\Big] +{\rm cyclic},
\end{eqnarray}
where $p_0=1$ and $p_m$ is the relative normalization of the polarization tensor of helicity $m$ to the higher helicity $s$.  

Having determined the general formulas for the contribution of the partial massless higher-spin  to the two- and three-point functions, it remains to specify the depth $n$. The latter can be specified from the requirement that the partial massless higher-spin field should be constant on super-Hubble scales, that is it should have scaling dimension $\Delta=0$. Using Eq. (\ref{aa}),  the mass of the partially massless field of depth $n$ is 
\begin{eqnarray}
m^2=H^2(n-1)(2s-n).
\end{eqnarray}
Hence, the condition $\Delta=0$ gives  $n=s-1$. In other words,  partially massless higher-spin fields with spin $s$ and  $\Delta=0$ have depth $n=s-1$
and polarizations $(-s,\cdots, -2,2,\cdots, s)$. Then, the contribution to the two- and three-point functions from the partially massless higher-spin fields are 
%
\be
\Big<\phi_{\vec{k}}\phi_{-\vec{k}}\Big>'=
\frac{c_\phi}{k^3} \bigg[ 1+\sum_{m=0}^{s-2}p^{\phi}_{m+1}\sin^{2(s-m)}(\hat{k}\cdot \hat n) \bigg ],
\ee
and
\be
\Big< \phi_{\vec{k}_1}\phi_{\vec{k}_2}\phi_{\vec{k}_3}\Big>'_{\rm av}= 
\frac{c_{\phi\phi\phi}}{k_1^3k_2^3}
\sum_{m=0}^{s-2} p^{\phi\phi\phi}_m\cos^m(\hat{k}_1\cdot \hat{k}_2)\Big[1+\cos^{2s-m}(\hat{k}_1\cdot \hat{k}_2)\Big]+{\rm cyclic}.
\ee
 As for the four-point correlator in the collapsed limit, the statistically anisotropic contribution from the partially massless states reads
  \be
 \label{asd}
\Big< \phi_{\vk_1}\phi_{\vk_2}\phi_{\vk_3}\phi_{\vk_4}\Big>^\prime \supset \frac{c_{\phi\phi\phi\phi}
}{k_{12}^3 k_2^3 k_4^3}\sum_{m=0}^{s-2} p^{\phi\phi\phi\phi}_m
\sin^{2(s-m)}(\hat k_{12}\cdot \hat n)+
\,\,{\rm cyclic},\,\,\,\,\,\,\,\,\,\,(\vk_{12}=\vk_1+\vk_2\simeq \vec{0}).
\ee
 We leave the relative coefficients $p_m$ among the various helicities free as in general one does not expect the respective  vacuum expectation value to be  related. If they are, the coefficients can be fixed using the special conformal transformations.

\section{Conclusions}
If light spinning particles exist during inflation, they leave a characteristic imprint on the scalar primordial cosmological perturbations. In this note we have investigated how the angle dependence of the statistical anisotropies should look like. We have been able to reproduce the well-known 
results of the spin-1 case and generalise it to higher-spin.

We conclude with a few  comments.
We have concentrated ourselves only on the calculation of the correlators
of the scalar perturbations. Among the various  correlators  we have computed, there is the one involving
higher-spin and scalar fluctuations. The significance of such
correlators is not clear since, differently from the case of the standard massless spin-2 graviton, the higher-spin states are most probably
doomed to decay and disappear after inflation. This will happen, for instance,  for those massive higher-spin states which are rendered effectively massless only during inflation.  
Furthermore, one could   investigate the possibility of the higher-spin field to play the role of the curvaton field and to be the ultimate responsible for the
curvature perturbation. 

 In the case of partially massless higher-spin fields  one will also require the construction of gauge-invariant quantities for which correlation functions will not probably diverge in the infrared and also give physical significance to the $\Delta=0$ states, possibly by constructing generalized curvature terms  for higher-spin fields \cite{generalized}. The same
 will be required for higher-spin fields which acquire vanishing  scaling dimension through couplings to matter. In such a case, as deduced from Ref. \cite{noi} these couplings
 will possibly involve mass terms an therefore they will not necessarily suffer of the same strong coupling problem of the corresponding spin-1 case. Also, if couplings to matter will
 take place through gauge-invariant quantities, one might worry about 
violating the necessary  constraint which  ensures that no ghost or extra states are propagating.  In fact, one might violate the basic property that the   constraints  are only first order in time derivatives of the fields.  In fact,  as  the  new possible contributions will only  involve dynamical matter fields, the unwanted  time derivatives can  be removed by making use of the matter field equations, see for instance Ref. \cite{gdeser}.

Finally, there exist consistent theories of higher-spin four-dimensional gravity  which come from the start with an infinite tower of massless higher-spin states  
 \cite{vas}. They have  been studied from the dS/CFT$_3$ correspondence point of view  \cite{stromingerhs} and very recently
  a consistent characterisation  of the Hilbert space of higher-spin  quantum gravity in de Sitter has been proposed
  Ref. \cite{anninos}.  Interactions among the higher-spin states can be studied on the 
 holographic dual side \cite{sthesis} and FRW-like solutions have been constructed in Ref.  \cite{norena}. It is interesting to note that higher-spin fields with $s\geq 2$ do show the infrared phenomenon of  accumulation of frozen modes and the consequent appearance of isotropy-breaking classical backgrounds.  Higher-spin interactions with the standard massless graviton
 can also be a new source of gravitational waves from inflation.
 
  These theories are  possibly related to tensionless string theories.  Strings have no tension and therefore their higher vibrational modes are unsuppressed, leading to an infinite tower of massless fields. Embedding this higher-spin theory into string theory and  deforming  the boundary field theory in a manner that turns on the bulk string tension will Higgs all of the higher-spin modes, giving them masses but leaving the bulk graviton massless \cite{higgs}. It will be interesting to see
if this phenomenon may deliver vanishing conformal weights and if it has some implication for inflation and its cosmological  observables.

\section*{Acknowledgments}
 A.K. is partially supported by GGET
project 71644/28.4.16. A.R. is supported by the Swiss National Science Foundation (SNSF), project {\sl Investigating the
Nature of Dark Matter}, project number: 200020-159223. We thank the authors of Ref. \cite{am1} for interesting discussions and for comments on our draft.

\appendix
\numberwithin{equation}{section}
%
%
%

\end{document}